\documentclass[lettersize,journal]{IEEEtran}
\IEEEoverridecommandlockouts
\usepackage{amsmath,amsfonts}
\usepackage{algorithmic}
\usepackage{algorithm}
\usepackage{array}
\usepackage[caption=false,font=normalsize,labelfont=sf,textfont=sf]{subfig}
\usepackage{textcomp}
\usepackage{stfloats}
\usepackage{url}
\usepackage{graphicx}
\usepackage{cite}
\usepackage{xcolor}
\usepackage[acronym]{glossaries}
\newacronym{3gpp}{3GPP}{3rd Generation Partnership Project}

\newacronym{ai}{AI}{Artificial Intelligence}
\newacronym{aoi}{AoI}{Age of Information}

\newacronym{bs}{BS}{Base Station}
\newacronym{cff_frame}{CFC-pull/push}{contention-free pull and contention-push}
\newacronym{cps}{CPS}{cyber-physical systems}
\newacronym{cowu}{CoWu}{Content-based Wake-up}
\newacronym{dl}{DL}{Deep Learning}
\newacronym{dt}{DT}{digital twin}
\newacronym{embb}{eMBB}{enhanced Mobile Broadband}
\newacronym{es}{ES}{Edge Server}

\newacronym{gnn}{GNN}{Graph Neural Network}
\newacronym{gNB}{gNB}{next-generation Node B}

\newacronym{iiot}{IIoT}{Industrial Internet of Things}
\newacronym{iot}{IoT}{Internet of Things}

\newacronym{kpi}{KPI}{Key Performance Indicator}
\newacronym{mac}{MAC}{Medium Access Control}
\newacronym{mcu}{MCU}{Micro Controller Unit}
\newacronym{ml}{ML}{Machine Learning}
\newacronym{mmtc}{mMTC}{massive Machine-type Communications}
\newacronym{mse}{MSE}{Meas Square Error}
\newacronym{nrt}{Near-RT}{Near-Real-Time}
\newacronym{near-rt-ric}{Near-RT RIC}{Near-Real-Time RAN Intelligent Controller}
\newacronym{non-rt-ric}{Non-RT RIC}{Non-Real-Time RIC}
\newacronym{o-ran}{O-RAN}{Open-Radio Access Network}

\newacronym{pdcch}{PDCCH}{Physical Downlink Control Channel}
\newacronym{phy}{PHY}{physical}
\newacronym{prach}{PRACH}{Physical Random Access Channel}
\newacronym{pusch}{PUSCH}{Physical Uplink Shared Channel}

\newacronym{qaoi}{QAoI}{Query Age of Information}

\newacronym{rcs_frame}{RCSC-pull/push}{Reserved pull-contention and shared pull-push contention}
\newacronym{rrc}{RRC}{Radio Resource Control}
\newacronym{ric}{RIC}{RAN Intelligent Controller}
\newacronym{semdas}{SEMDAS}{Semantic Data Sourcing}
\newacronym{snn}{SNNs}{Spiking Neural Networks}

\newacronym{twi}{TWI}{Temporal Windows of Integration}
\newacronym{ucwu}{UCWu}{unicast wake-up}
\newacronym{urllc}{URLLC}{Ultra-Reliable and Low Latency Communications}

\newacronym{voi}{VoI}{Value of Information}
\newacronym{wsn}{WSN}{Wireless Sensor Network}
\newacronym{wur}{WuR}{Wake-up Radio}
\newacronym{wus}{WuS}{Wake-up Signal}

\newacronym{xr}{XR}{eXtended Reality}

\usepackage[normalem]{ulem}

\usepackage{tikz}
\usepackage{tikzscale}
\usepackage{pgfplots}
\usepackage{pgfplotstable}
\pgfplotsset{
    compat=1.3,
    legend style={font=\scriptsize, fill opacity=1,  draw opacity=1, text opacity=1, draw=white!15!black, legend cell align=left, align=left},     
    ymajorgrids=true,
    xmajorgrids=true,    
    yminorticks=false,
    xminorticks=false,
    grid style={dashed},
    title style={font=\small},
    label style={font=\footnotesize},
    tick label style={font=\footnotesize},    
    tick align=inside,
    axis background/.style={fill=white},
    ylabel shift=-3pt,
}
\usepgfplotslibrary{fillbetween}
\usetikzlibrary{patterns,arrows,plotmarks}
\usepgfplotslibrary{groupplots}
\pgfdeclarelayer{background}
\pgfsetlayers{background,main}
\usetikzlibrary{automata,positioning}
\usetikzlibrary{decorations}
\usetikzlibrary{shapes.arrows}
\usetikzlibrary{tikzmark}
\usetikzlibrary{calc}
\usetikzlibrary{decorations.markings}
\usepgfplotslibrary{colorbrewer}

\definecolor{amaranth}{rgb}{0.9, 0.17, 0.31}
\definecolor{steelblue}{RGB}{176,196,222}
\definecolor{darkblue}{RGB}{0,0,139}
\definecolor{lightblue}{RGB}{31,119,180}
\definecolor{deepskyblue}{RGB}{0,191,255}
\definecolor{lightskyblue}{RGB}{135,206,250}
\definecolor{lightgray}{rgb}{0.82, 0.82, 0.82}
\definecolor{gray}{RGB}{140,140,140}
\definecolor{darkgray204}{RGB}{204,204,204}
\definecolor{darkgray224}{RGB}{224,224,224}

\newcommand{\rv}[1]{{#1}}
\newcommand{\minor}[1]{\textcolor{black}{#1}}

\begin{document}

\title{Medium Access for Push-Pull Data Transmission \\ in 6G Wireless Systems}

\author{Shashi Raj Pandey,~\IEEEmembership{Member,~IEEE,}  Fabio Saggese,~\IEEEmembership{Member,~IEEE,} Junya Shiraishi,~\IEEEmembership{Member,~IEEE,} \\Federico Chiariotti,~\IEEEmembership{Senior Member,~IEEE}, Petar Popovski,~\IEEEmembership{Fellow,~IEEE}
\thanks{S. R. Pandey (srp@es.aau.dk), J. Shiraishi (jush@es.aau.dk), and P. Popovski (petarp@es.aau.dk) are with the Electronic Systems Dep., Aalborg University, Denmark. F. Saggese is with Dept. of Information Engineering, University of Pisa, Italy (fabio.saggese@ing.unipi.it), and his work is funded by Horizon Europe MSCA Postdoctoral Fellowships with Grant~101204088. F. Chiariotti (federico.chiariotti@unipd.it) is with the Department of Information Engineering, University of Padova, Italy. This work was partly supported by the Villum Investigator Grant ``WATER" from the Velux Foundation, Denmark, partly by the Horizon Europe SNS ``6G-XCEL" project with Grant 101139194, and partly by the Horizon Europe SNS ``6G-GOALS'' project with grant 101139232. The work of S.R. Pandey was supported in parts by DFF-Forskningsprojekt1 with grant No. 4286-00278B. The work of J. Shiraishi was supported by Horizon Europe MSCA Postdoctoral Fellowships ``NEUTRINAI" with grant No.~101151067.
\minor{The simulation codes are available at https://github.com/lostinafro/pushpull-mag}
}}



\maketitle

\begin{abstract}
Medium access in 5G systems was tailored to accommodate diverse traffic classes through network resource slicing. 
6G wireless systems are expected to be significantly reliant on Artificial Intelligence (AI), leading to data-driven and goal-oriented communication. This leads to augmentation of the design space for Medium Access Control (MAC) protocols, which is the focus of this article. We introduce a taxonomy based on push-based and pull-based communication, which is useful to categorize both the legacy and the AI-driven access schemes. We provide MAC protocol design guidelines for pull- and push-based communication in terms of goal-oriented criteria, such as timing and data relevance. We articulate a framework for co-existence between pull and push-based communications in 6G systems, combining their advantages. We highlight the design principles and main tradeoffs, as well as the architectural considerations for integrating these designs in Open-Radio Access Network (O-RAN) and 6G systems.
\end{abstract}

\begin{IEEEkeywords}
IoT connectivity, pull-based communication, push-based communication, 
medium access control, timing, data relevance 
\end{IEEEkeywords}

\section{Introduction}
One of the major promises of the sixth generation of mobile networks (6G) is a seamless integration of technology, data, and intelligence in real time~\cite{wang2023road}. This is a key enabler for new applications such as \gls{xr} and \glspl{dt}, which require a two-way connection between the digital and physical worlds: decisions in the digital domain have real, tangible effects, and real events need to be integrated in the digital model of the environment. Naturally, this level of real-time integration may strain even the capacity of future 6G networks, particularly when there is a massive number of connected devices\rv{~\cite{lopez2023energy}} and the computational requirements of \gls{ai} models are considered.

The novel services envisioned with 6G systems expect flexible communication protocols designed to accommodate different applications and requirements. The \emph{network slicing} approaches proposed for 5G can create multiple virtual networks tailored to specific service needs on a shared physical infrastructure to meet the requirements of different traffic classes. 
However, due to the dynamic variability of end-to-end goals over time as well as the use of AI by the communicating parties, 6G systems will require agile and robust \gls{mac} protocols to accommodate diverse application-driven requirements, not catered efficiently by the existing network slicing strategies~\cite{wang2023road}. \minor{As such, the \gls{mac} protocols should, by design, adhere to goal-oriented communication approaches, which is missing in the related works on protocols.} The communication protocols for 6G systems should be designed considering the growing intelligence in the communicating nodes and the coexistence of diverse application requirements and communication modalities.

The resulting high reliability and flexibility requirements pose two key challenges: first, \textit{where} to process the information under strict latency-reliability requirements, and secondly, \textit{when} and \textit{how} to collect the data for processing. The former aspect has been extensively discussed in the domain of edge computing and task offloading. The latter aligns with the goal-oriented communication paradigm~\cite{gunduz2022beyond}, which optimizes communication decisions based on the \gls{voi} of updates: when transmitting all available data is impossible due to timing or energy constraints, goal-oriented schemes privilege information that is both \textit{novel} and \textit{relevant} for the receiver, i.e., is both not directly predictable from previously available information and useful for the task at hand\rv{~\cite{talli2025pragmatic,Kountouris2021semantic}}.

The increased role of AI gave rise to the notions of semantic and goal-oriented communication, focusing on the extraction of relevant information~\cite{gunduz2022beyond}. Goal-oriented optimization of \gls{mac} mechanisms entails coordination of multiple agents with a partial view of the system. For example, uncoordinated decisions made by \gls{iot} devices for uplink transmission are myopic and may deteriorate performance due to the risk of collisions. The majority of existing goal-oriented schemes operate in a \textit{pull-based} fashion~\cite{akar2024query}: the \gls{bs} is directly connected to the \gls{dt} of the system and is thus able to estimate which information is expected to be more relevant, proactively requesting it from the relevant devices. However, pull-based systems can only rely on statistical information, while the sensors themselves have access to the actual observed information and may have a more precise idea of its value. Ideally, decentralized \textit{push-based} access may lead to more informative updates and a better use of wireless resources~\cite{talli2025pragmatic}. However, this brings challenges of coordination, as individual sensors have a limited view of the environment, and random access schemes have well-known congestion issues.

\begin{figure*}[t!]
    \centering
    \includegraphics[width=0.9\textwidth]{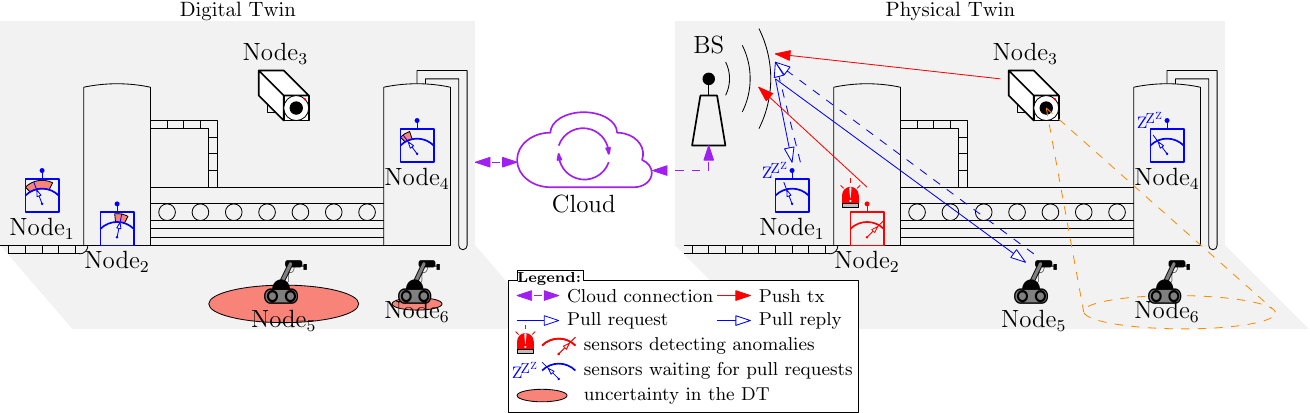}
    \caption{A toy example of an \gls{iot} scenario where a \gls{bs} collects measurements from \gls{iot} devices to construct a \gls{dt} of the system. The coexistence of push- and pull-based communication enables the \gls{bs} to coordinate requests for the most informative data to update the \gls{dt} model while allowing \gls{iot} devices to immediately report detected anomalies.}
    \label{fig:cps}
\end{figure*}

In fact, the two approaches have complementary benefits: push-based communication follows an \emph{event-driven} approach, excelling in scenarios that require immediate reporting and response, such as anomaly detection. Pull-based communication is more effective in scenarios where the edge or cloud runs a comprehensive AI model that can impose global orchestration of the data collection and selectively retrieve data, thereby reducing irrelevant data traffic. Moreover, it offers flexibility in adopting communication strategies to handle requests for different tasks by, for example, deciding the timing of data collection to achieve a specific goal~\cite{ildiz2023pull}. This is evident in the scenario shown in Fig.~\ref{fig:cps}: the uncertainty of the \gls{dt} model over the robot (node 5) is high, so the \gls{bs} can initiate a pull request to get that information. In this case, the global view of the system allows the \gls{bs} to reduce communication by, e.g., exploiting the information from the camera (node 3) to avoid contacting the other robot (node 6) as well. On the other hand, one of the sensors (node 2) is registering an anomaly, which is not reflected in the \gls{dt}. This event would be detected very late under a purely pull-based scheme, and requires the node to be able to transmit at will.

\begin{figure}
    \centering
    \includegraphics[width=\columnwidth]{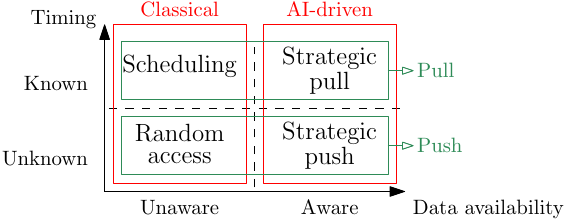}
    \caption{Design space of push and pull communication paradigms depending on the data availability and timing of arrival.}
    \label{fig:taxonomy}
\end{figure}

The dichotomy between classical and goal-oriented schemes then needs to be complemented by a second axis, namely, push- versus pull-based schemes. This distinction is already present in classical schemes, as scheduled mechanisms are pull-based, while random access is push-based. Fig.~\ref{fig:taxonomy} shows a graphical representation of these aspects: while pull-based schemes rely on centralized knowledge of packet timing, which push-based schemes lack, the distinction between classical and AI-driven semantic schemes~\cite{shi2021semantic} hinges on the availability of data, i.e., knowledge about the meaning and value of data that each sensor has, and the awareness of timing for data availability. This adds a novel semantic dimension in the design space of access schemes, as interactions between the \gls{bs} and the devices are \emph{strategic}. While classical \gls{mac} schemes operate independently from this aspect, it is at the center of goal-oriented communication schemes, which target the \gls{voi} of potential updates, privileging more informative updates for the \gls{dt} of the environment. 

The design of protocols and \gls{mac} mechanisms for the coexistence of push- and pull-based traffic supporting goal-oriented \glspl{kpi}, such as estimation error or control performance, in addition to traditional timing metrics, is a key enabler for real-time applications such as \glspl{dt} and \gls{xr}~\cite{thomas2023causal}; addressing this gap is imperative to the design of efficient 6G systems. As knowledge about the environment is itself distributed, centralizing decision-making at the \gls{bs} or distributing it to the sensors leads to highly different information structures and signaling requirements. Our previous work~\cite{cavallero2024co-existence,shiraishi2024co-existence} has begun exploring some of these challenges, but this paper provides a systematic treatment of integrated access framework, considering the fact that AI-driven operation expands the design space of medium access protocols.

This article introduces the design criteria for \minor{future} \gls{mac} protocols to accommodate push- and pull-based communications, offering guidelines for organizing intelligent operation in 6G wireless access networks, as explained in sections~\ref{sec:pull_and_push} and~\ref{sec:protocol_design}. \minor{To this end, we provide a time-division frame structure that can accommodate these novel schemes, as well as analyzing} their potential for application-driven 6G network operation in section~\ref{sec:integration_challenges}.

\section{Goal-Oriented Communication: A MAC View}
\label{sec:pull_and_push}

In a goal-oriented scenario such as the one depicted in Fig.~\ref{fig:cps}, each sensor may have precise knowledge of specific physical parameters based on its own measurements, but this knowledge is limited to the status of a local part of the environment: the sensor has no way to obtain a \textit{global} view of the environment, other than through a transmission from the \gls{bs}. Aside from the signaling issues required to maintain and synchronize even a partial copy of the environment \gls{dt} at each sensor, there are also significant computational limitations: sensors are often simple battery-powered devices with low-power computing units, which cannot run \gls{ai} models in real time. Pull-based communication seems to be the best possible solution to this issue~\cite{ildiz2023pull}; since the \gls{bs} is the only node with a global view of the system, centralizing decisions is an attractive option. Centralized scheduling also mitigates the collision problem, as it allows for orthogonal division of channel resources to avoid interference. 

On the other hand, having a global view and easily coordinating access comes at a cost: since it is based on a statistical model, it relies on the expected behavior of the system, and it might not reflect physical reality. 
Unexpected events, such as anomalies or component failures, might not be accurately modeled, or assigned a very low probability. Only push-based communication, in which individual sensors decide when to transmit based on the actual measurements they make~\cite{talli2025pragmatic}, is able to overcome this issue and quickly make the \gls{bs} aware of the anomaly. However, push-based schemes suffer from coordination problems, negating the advantages of pull-based scheduling. 

More complex sampling schemes, such as \gls{semdas}~\cite{huang2023semantic}, allow for content-based or semantic queries, which direct sensors to transmit if their measurements are within the (possibly complex) parameters of the query. This hybrid solution maintains and even enhances the advantages of having a global view of the system, but loses the coordination advantages of purely pull-based solutions, as the number and identities of the selected nodes might be unknown even to the \gls{bs} itself. However, the fundamental trade-off remains, and the relevant literature is still considering push- and pull-based schemes as mutually exclusive. The coexistence between push- and pull-based access in the same system, enabling both global coordination for normal operation and quick anomaly reporting, is a new challenge that opens new fundamental trade-offs and design problems.

\label{sec:Protocol_design}
\begin{figure*}[t!]
    \centering   
    \includegraphics[width=.9\textwidth]{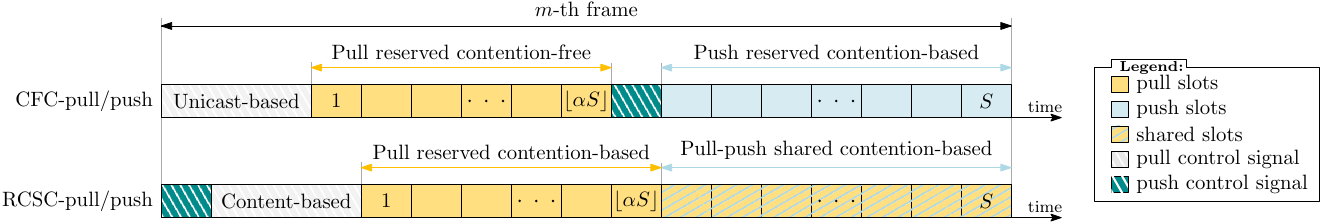}
    \caption{A time-diagram for the considered push-pull frame structures.}
    \label{fig:mac-protocol}
\end{figure*}
Deciding whether to allocate specific resources to push-based operation, and how to balance between pull- and push-based transmission, represents a complex problem that depends on the application requirements. Static, predefined allocation schemes cannot adapt to the conditions of the environment, while dynamic schemes require additional signaling to inform nodes of the resource allocation. Moreover, coordinating push-based transmissions becomes even more complex if nodes are also regularly transmitting data in a pull-based fashion, changing the push-based decision-making due to recent or upcoming pull-based transmission opportunities.
\section{Protocol design}\label{sec:protocol_design}
Combining the push- and pull-based paradigms requires careful protocol designs, accounting for application-level requirements and \glspl{kpi}. In this section, we highlight the design principles for communication protocols managing the coexistence of push- and pull-based data transmission schemes, along with key trade-offs for two specific goal-oriented scenarios: centralized decision-making and decentralized intelligence. 

\subsection{Centralized decision-making}
\label{sec:Protocol_centralized}
In the centralized decision-making scenario, the \gls{bs} controls medium access for both push- and pull-based traffic considering the application-level requirement and expected traffic for each class~\cite{cavallero2024co-existence}. Our recent work outlines two types of \gls{mac} layer designs, as illustrated in Fig.~\ref{fig:mac-protocol}. One is the \emph{\gls{cff_frame}}, which consists of two parts: scheduled contention-free slots for pull-based communication and random access contention-based slots for push-based communication~\cite{cavallero2024co-existence}. The other is \emph{\gls{rcs_frame}}, which also consists of two parts: reserved random access periods for pull-based traffic and shared random access periods for both traffic classes~\cite{shiraishi2024co-existence}. In both structures, the total available time slots per frame are organized into corresponding sub-frames by adjusting a control parameter $\alpha\in[0,1]$, representing the fraction of resources reserved for pull-based communication. 

\begin{figure}
    \centering
    \begin{tikzpicture}

\def\vside{3.2cm}

\def\XshiftA{0pt}
\def\YshiftA{0pt}
\def\XshiftB{-0.8pt}
\def\YshiftB{-1pt}

\begin{axis}[
    height=\vside,
    width=.8\columnwidth,
    scale only axis,
    legend cell align={left},
    legend style={  
      at={(0.5, 1.0)}, 
      draw=none,
      fill opacity=0,
      anchor=south, 
      /tikz/every even column/.append style={column sep=0.2cm},
      legend columns=-1,
    },
    xmajorgrids,    
    xlabel shift=-3pt,
    xlabel={Avg. achievable incoming pull traffic [k-packets/s]},
    xmin=0, xmax=3000,
    xtick={0,500,1000,1500,2000,2500,3000},
    xticklabels={0,0.5,1,1.5,2,2.5,3},
    ylabel shift=-4pt,
    ylabel style={text width=\vside, align=center},
    ylabel={Avg. achievable incoming push traffic [k-packets/s]},
    ymajorgrids,
    ymin=0, ymax=5300,
    ytick={0,1000,2000,3000,4000,5000},
    yticklabels={0,1,2,3,4,5},
    point meta=explicit symbolic,    
    nodes near coords, 
    nodes near coords style={
        text=black,
        draw=none,
        anchor=south,
        font=\tiny,
    },
]
\addlegendimage{draw=gray, fill=white, mark=*, only marks, mark size=1.2}
\addlegendentry{$\alpha$ ($\%$)}
\addlegendimage{draw=gray}
\addlegendentry{$L = 20$~ms}
\addlegendimage{draw=darkblue}
\addlegendentry{$L = 30$~ms}
\addlegendimage{draw=deepskyblue}
\addlegendentry{$L = 50$~ms}

\addplot [draw=gray, fill=gray, mark=*, mark size=1.2pt, only marks, forget plot, nodes near coords style={yshift=-0.7mm}]
table[x=pull, y=push, meta=Q]{tabletwo.dat};
\addplot[semithick, gray, const plot mark right, forget plot]
table[x=pull, y=push]{tabletwo.dat};

\addplot [draw=darkblue, fill=darkblue, mark=*, mark size=1.2pt, only marks, forget plot, nodes near coords style={yshift=-0.7mm}]
table[x=pull, y=push, meta=Q]{tablethree.dat};
\addplot[semithick, darkblue, const plot mark right, forget plot]
table[x=pull, y=push]{tablethree.dat};

\addplot [draw=deepskyblue, fill=deepskyblue, mark=*, mark size=1.2pt, only marks, forget plot, nodes near coords style={yshift=-0.7mm}]
table[x=pull, y=push, meta=Q]{tablefive.dat};
\addplot[semithick, deepskyblue, const plot mark right, forget plot]
table[x=pull, y=push]{tablefive.dat};
\end{axis}

\end{tikzpicture}
    \caption{\gls{cff_frame} achievable incoming traffic satisfying a target latency $L$ with reliability 99\% as a function of $\alpha$~\cite{cavallero2024co-existence}.}
    \label{fig:cfc-result}
\end{figure}
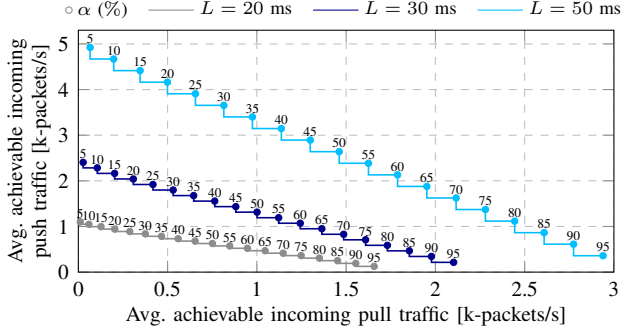

\gls{cff_frame} is a frame structure similar to the contention-free and contention-based mechanism in WiFi technology. Each frame's organization is notified to the devices through control signaling~\cite{cavallero2024co-existence}: firstly, the \gls{bs} decides the \emph{device scheduling} for the pull-based communication phase, taking into account the available pull-slots, and notifies devices by broadcasting the scheduling information. This \emph{unicast scheduling} enables the \gls{bs} to retrieve sensor readings from specific devices at specific times without packet collisions. 
Conversely, during the push-contention phase, each device transmits an update -- if it deems it is important enough -- by contending for the channel with other push-based traffic. The major challenge is to design the resource allocation scheme that maximizes the goal-oriented utility for both phases~\cite{cavallero2024co-existence}. 
Fig.~\ref{fig:cfc-result} shows the maximum average incoming traffic that can be handled by the system when constraining the successful delivery of packets within a target latency $L$ with reliability $99\%$. \minor{Simulations are made according to the following setting: a 10 ms frame comprises a total of $S = 100$ slots, with pull and push packets occupying 5 and 1 slots each, respectively, due to the different data carried~\cite{cavallero2024co-existence}; a 1-persistent retransmission scheme in case of push collision is employed; similarly, unscheduled pull packets are rescheduled in the next frame.} Every point in the plot represents a different percentage of the slots reserved for pull requests, $\alpha$. Any value above the curves cannot satisfy the latency constraint, giving a measure of scalability of the system with respect to the incoming system load. As expected, increasing this value leads to a decrease in the acceptable push traffic, while accommodating a higher pull traffic, and vice versa. Naturally, reducing the target latency allows the \gls{bs} to manage a higher overall load.

\begin{figure}
    \centering
    \begin{tikzpicture}

\definecolor{crimson2143940}{RGB}{214,39,40}
\definecolor{darkgray176}{RGB}{176,176,176}
\definecolor{darkorange25512714}{RGB}{255,127,14}
\definecolor{forestgreen4416044}{RGB}{44,160,44}
\definecolor{lightgray204}{RGB}{204,204,204}
\definecolor{mediumpurple148103189}{RGB}{148,103,189}

\def\hsep{1.2cm}
\def\vside{3.2cm}
\def\hside{0.65\columnwidth}

\pgfplotsset{set layers}

\begin{axis}[
height=\vside,
width=\hside,
scale only axis,
xmajorgrids,
xlabel={Portion of pull reserved slots, $\alpha$},
xmin=0, xmax=1,
ymin=0, ymax=1,
axis y line*=left,
ytick={0,0.25,0.5,0.75,1},
ymajorgrids,
yticklabels={
  \(\displaystyle {0.0}\),
  \(\displaystyle {0.25}\),
  \(\displaystyle {0.50}\),
  \(\displaystyle {0.75}\),
  \(\displaystyle {1.0}\),
},
ylabel shift=-4pt,
xlabel shift=-3pt,
ylabel style={text width=\vside, align=center},
ylabel={Pull retrieval accuracy \\ (solid lines)}
]
\addplot [thick, orange, forget plot, mark=asterisk, mark size=1.5, mark options={solid, fill=white}]
table {pull_s25.dat};

\addplot [thick, cyan, forget plot, mark=*, mark size=1.5, mark options={solid, fill=white}]
table {pull_s50.dat};

\addplot [thick, red, forget plot, mark=square*, mark size=1.5, mark options={solid, fill=white}]
table {pull_s75.dat};
\end{axis}

\begin{axis}[
height=\vside,
width=\hside,
scale only axis,
legend style={  
  at={(0.5, 1.00)}, 
  draw=none,
  fill opacity=0,
  anchor=south,  
  /tikz/every even column/.append style={column sep=0.2cm}
},
legend columns=-1,
xmajorgrids,
xmin=0, xmax=1,
ymin=0, ymax=1,
axis y line*=right,
axis x line=none,
ytick={0,0.25,0.5,0.75,1},
ymajorgrids,
yticklabels={
  \(\displaystyle {0.0}\),
  \(\displaystyle {0.25}\),
  \(\displaystyle {0.50}\),
  \(\displaystyle {0.75}\),
  \(\displaystyle {1.0}\),
},
ylabel shift=-4pt,
ylabel style={align=center},
ylabel={Push success probability \\ (dashed lines)}
]
\addplot [thick, orange, dashed, mark=asterisk, mark size=1.5, mark options={solid, fill=white}, forget plot]
table {push_s25.dat};

\addplot [thick, cyan, dashed, mark=*, mark size=1.5, mark options={solid, fill=white}, forget plot]
table {push_s50.dat};

\addplot [thick, red, dashed, mark=square*, mark size=1.5, mark options={solid, fill=white}, forget plot]
table {push_s75.dat};

\addplot [thick, orange, mark=asterisk, mark size=1.5, mark options={solid, fill=white}, only marks]
table {marker_s25.dat};
\addlegendentry{$S=25$}

\addplot [thick, cyan, mark=*, mark size=1.5, mark options={solid, fill=white}, only marks]
table {marker_s50.dat};
\addlegendentry{$S=50$}

\addplot [thick, red, mark=square*, mark size=1.5, mark options={solid, fill=white}, only marks]
table {marker_s75.dat};
\addlegendentry{$S=75$}
\end{axis}




\end{tikzpicture}
    \caption{\gls{rcs_frame} performance in terms of pull retrieval accuracy (solid lines) and push success probability (dashed lines) as a function of $\alpha$, and for different number $S$ of slots in a frame.}
    \label{fig:rcsc-results}
\end{figure}
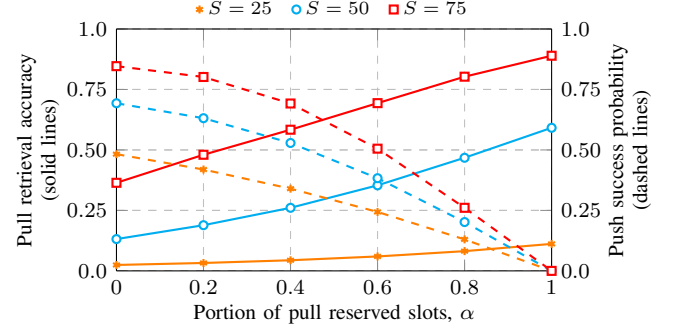

Protocol designs based on \gls{rcs_frame} are crucial when data collection by the \gls{bs} is based on the specific observations of each device~\cite{shiraishi2024co-existence}. The \gls{rcs_frame} can be divided into two portions: pull-reserved contention slots and pull-push shared contention slots. At the beginning of the frame, the \gls{bs} sends a control signal to push-enabled devices to notify them of the frame structure. The \gls{bs} then broadcasts a semantic query, which specifies the features of the data that the \gls{bs} considers relevant~\cite{shiraishi2024co-existence}, e.g., by specifying a range over the value of sensor observations, to all pull-enabled devices. Pull-enabled devices receiving this query check whether their observations satisfy the condition specified by the \gls{bs} and transmit based on a random access scheme. On the other hand, devices operating in a push-based fashion start transmitting data at the beginning of the shared slots if they observe informative data, even if it is outside the query specifications. 
Within the shared slots, push-based transmissions need to contend for the channel with the pull-based responses that have not yet been successful by the end of the pull reserved slots. This results in both intra-class and inter-class collisions~\cite{shiraishi2024co-existence}. Optimizing this frame structure entails trade-offs between different task-specific communication requirements. Fig.~\ref{fig:rcsc-results} illustrates the impact of varying $\alpha$ on the content-driven communication scenario. Performance is evaluated in terms of the pull-based retrieval accuracy and push-based transmission success probability. The former represents the probability that the \gls{bs} successfully receives all requested content, while the latter is the transmission success probability for pushed data. Fig.~\ref{fig:rcsc-results} shows that increasing $\alpha$ improves pull-based retrieval accuracy, since most sensors are able to successfully transmit in the reserved slots. On the other hand, higher values of $\alpha$ reduce the push-based success probability due to fewer available transmission opportunities and more intra-class collisions.

\minor{The proposed frame designs are general and allow the use of any scheduler and random access approach tailored for managing transmissions within contention-free and contention-based slots, respectively. More advanced contention resolution techniques -- e.g., by making the use of successive interference cancellation -- can be integrated to enhance the access performance in contention-based sub-frames, though the quantification of potential gains requires further investigation.}

\subsection{Decentralized intelligence}\label{sec:Distributed intelligence}
In the decentralized intelligence scenario, devices make local observations and estimate their expected value to decide when to transmit. As individual devices have limited knowledge about others' observations, transmissions are subject to collisions, and thus to the loss of timely and accurate status updates.
Devices then need to decide whether and when to attempt a transmission so as to satisfy the \glspl{kpi} for the goal-oriented tasks, such as \gls{voi}~\cite{talli2025pragmatic}. As even the \gls{bs} has partial information on the value of data at the nodes, the access protocols must adapt to accommodate push- and pull-based traffic depending on the requirements set by the emerging applications and use cases, exploiting the flexibility in the proposed frame structure. Additionally, devices might also have limited computational and storage capabilities: in these settings, \minor{the decision-making needs} to be shifted towards the \gls{bs}, but centrally computed policies that can be executed locally may still provide good performance.

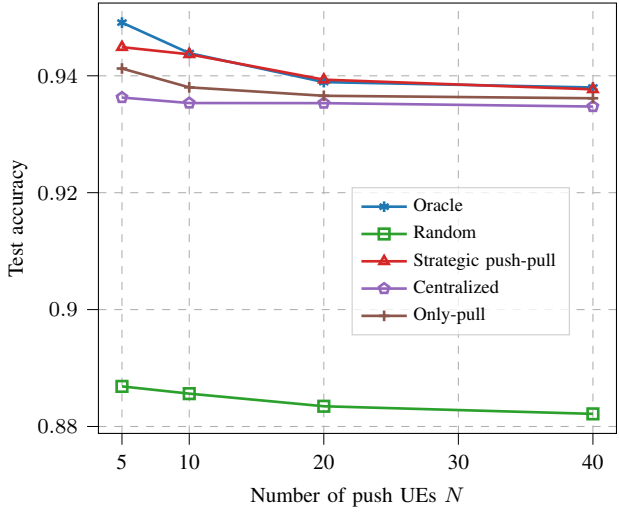
\begin{figure}[t!]
\centering
\begin{tikzpicture}

\definecolor{crimson2143940}{RGB}{214,39,40}
\definecolor{darkgray176}{RGB}{176,176,176}
\definecolor{darkorange25512714}{RGB}{255,127,14}
\definecolor{forestgreen4416044}{RGB}{44,160,44}
\definecolor{lightgray204}{RGB}{204,204,204}
\definecolor{mediumpurple148103189}{RGB}{148,103,189}
\definecolor{sienna1408675}{RGB}{140,86,75}
\definecolor{steelblue31119180}{RGB}{31,119,180}

\def\hsep{1.2cm}
\def\vside{3.2cm}
\def\hside{0.65\columnwidth}

\pgfplotsset{set layers}
\begin{axis}[
legend cell align={left},
legend style={
  fill opacity=0.8,
  draw opacity=1,
  text opacity=1,
  at={(0.91,0.4)},
  anchor=east,
  draw=lightgray204
},
tick align=outside,
tick pos=left,
xtick={50,100, 200, 300, 400},
xticklabels={5,10, 20, 30, 40},
x grid style={darkgray176},
xlabel={Number of push UEs $N$},
xmin=32.5, xmax=417.5,
xtick style={color=black},
y grid style={darkgray176},
ylabel={Test accuracy },
ymin=0.878813001811504, ymax=0.952447005808354,
ytick style={color=black}
]
\addplot [semithick, steelblue31119180, mark=asterisk, mark size=2, mark options={solid}, line width =1]
table {oracle.dat};
\addlegendentry{Oracle}

\addplot [semithick, forestgreen4416044, mark=square, mark size=2, mark options={solid}, line width =1]
table {random.dat};
\addlegendentry{Random}

\addplot [semithick, crimson2143940, mark=triangle, mark size=2, mark options={solid}, line width =1]
table {strategic.dat};
\addlegendentry{Strategic push-pull}

\addplot [semithick, mediumpurple148103189, mark=pentagon, mark size=2, mark options={solid}, line width =1]
table {centralized.dat};
\addlegendentry{Centralized}

\addplot [semithick, sienna1408675, mark=+, mark size=2, mark options={solid}, line width =1]
table {onlypull.dat};
\addlegendentry{Only-pull}
\end{axis}

\end{tikzpicture}
\caption{Learning performance under a fixed number of slots per frame; strategic access push-pull improves accuracy over Only-pull and Centralized approach, where contention-free collection of fraction of data is done. The method stays competitive with the Oracle and outperforms Random scheduling significantly.}  
\label{fig:push-pull-learning}
\end{figure}

For instance, consider a distributed learning setting for collective intelligence. The \gls{bs} would use private data available at the \gls{iot} devices, which have limited computational and communication capabilities, to train high-quality learning models. To do so, the \gls{bs} can reuse the \gls{cff_frame} frame structure and the corresponding communication protocol to solicit application-relevant data. This means devising efficient scheduling policies to accelerate high-quality training, where contention-free pull-reserved slots are allocated to the devices with the most useful information in each frame. A wrong attribution of the value of data from specific devices during scheduling, due to the partial knowledge of the \gls{bs}, can result in a weaker model with poor generalization performance. To overcome this issue, devices that are not accommodated in the pull slots need \minor{(push)} transmission opportunities to \minor{autonomously} share a better representation of their data. This can be done by arranging available communication resources for \minor{both} pull- and push-based slots, such that a high-quality learning model can be trained in a communication-efficient manner, as shown in Fig.~\ref{fig:push-pull-learning}. The figure demonstrates how proposed strategic push-pull approach outperforms Random, Centralized, and Only-pull approaches in terms of test accuracy, while providing competitive accuracy to the Oracle scheme, where the full information on data availability and timing is exploited. A reference protocol design for such a case is defined in~\cite{bui2024time}. 

Furthermore, devices need to balance the urgency of updates that might have to wait for a long time before the next pull-based slot with the relative scarcity of push-based slots and the consequent risk of collisions. Goal-oriented operation might require \minor{inferences} about other devices' beliefs, exploiting publicly available information to come to a consensus without further signaling. Naturally, aside from the computational load that this puts on the nodes, it also requires a careful design of the beacon and acknowledgment messages. However, a full analysis of the additional energy consumption this imposes on the devices is strongly dependent on the mechanism or \gls{ai} tool that they use to estimate value and make decisions, and is beyond the scope of this paper.

\section{Integration Challenges and Opportunities}\label{sec:integration_challenges}
In this section, we identify and discuss key challenges in protocol designs to enable the integration of push-pull data transmission in 6G systems.

\subsection{Intelligent MAC protocols}
\label{sec:intelligent_mac}
The goal-oriented aspect of communication outlines the value of the data collected at the receiver, instead of quantifying it locally at the device collecting the data, following task-specific requirements. Therefore, the underlying \gls{mac} mechanism requires \emph{adaptability} and \emph{flexibility} to allow intelligent coordination between \gls{iot} devices and the \gls{bs}. As individual devices are oblivious to the global view, they can only act according to their local knowledge of the value of available data upon transmission when employing conventional push-based schemes. Alternatively, devices may be able to perform a \emph{strategic push} by inferring the potential value of data transmission through past feedback from the \gls{bs}, which implicitly reflects the knowledge on the state of the system. This can alleviate the complexity of distributed coordination and intelligence when optimizing \gls{mac}. Similarly, the \gls{bs} can exploit its knowledge of prior interactions and employ a more conservative mechanism in the following frames, performing a \emph{strategic pull}. 

However, the risk of missing out unforeseen critical updates, such as alarms, always exists in the absence of available push communication resources. In practice, the previous considerations require the \gls{bs} to be able to dynamically modify the \gls{mac} structure to achieve different trade-offs, switching between \gls{cff_frame} and \gls{rcs_frame} frames and adjusting the number of pull, push, or shared slots (see Fig.~\ref{fig:mac-protocol}). This also translates to the need for specific and reliable control signals broadcasting the information regarding the \gls{mac} design to all devices in a timely manner. In the extreme case, the \gls{mac} structure could be modified on a frame-by-frame basis, but trade-offs between frame flexibility and the overhead caused by the additional control information should be investigated when considering the system-level performance of grant-based and grant-free access.

\subsection{Energy-Efficient Connectivity}\label{sec:sustainable_challenges}
Energy-efficient protocol designs for centralized decision-making and decentralized intelligence are crucial to realize energy-efficient connectivity in 6G networks. Technological enablers such as \gls{wur} can be leveraged to design sustainable communication systems and protocols that support intelligent tasks at the network edge~\cite{lopez2023energy}. Our recent work~\cite{cavallero2024co-existence, shiraishi2024co-existence} outlines communication framework designs incorporating \gls{wur} to manage push-pull coexistence, allowing \gls{iot} devices to keep their primary transceivers off until they detect a pull request. Specifically, \gls{ucwu}~\cite{piyare2017ultra} is introduced for \gls{cff_frame} to retrieve data from a specific device in a grant-based manner. Conversely, \gls{cowu} is introduced to enable content-based pull with \glspl{wur} in the \gls{rcs_frame} setting~\cite{shiraishi2024co-existence}. In \gls{cowu}, the information about the desired content is embedded into the \gls{wus}, allowing only the nodes with relevant data to activate their primary transceiver.

While employing \gls{wur} can improve the overall energy efficiency, it also adds design challenges. Highly informative feedback from the \gls{bs}, e.g., aiming to improve future transmission policy through strategic push, cannot be realistically performed through \glspl{wus} due to their low data rate~\cite{piyare2017ultra}. Furthermore, a dynamic adoption of \glspl{wus} within the flexible \gls{mac} protocols is needed to tailor the correct push-pull paradigm per the design space, accordingly to the current context of data collection and the goal-oriented scenario (see Fig.~\ref{fig:taxonomy}). An energy efficient implementation of such dynamic \gls{mac} reconfiguration requires a thorough investigation of the energy costs associated with signaling overhead. 

Compute operations needed for \gls{ai} training and inference add a significant energy load to the \gls{iot} networks. In a collaborative learning scenario, the communication overhead for distributed coordination between devices to collect/exchange model updates is significant. With the increasing importance of \gls{ai} technology in 6G networks, communication protocols should also take into account the energy cost of employing \gls{ai} models and their interaction with the communication system for goal-oriented tasks. Low-power \gls{ai} models, such as TinyML and \gls{snn}, can serve as technology enablers for understanding data semantics \emph{locally} and transmitting valuable model updates that improve the learning performance \cite{bui2024time}: the \gls{bs} can use the estimates of data availability, timing, and relevance for scheduling devices' updates during training, while allowing devices to also share learning models based on their local knowledge.

Specifically, while a \gls{bs} could retrieve only high-quality data that improves its model accuracy through strategic pull, the devices can strategically push relevant model parameters and data, or even anomalous samples detected locally by the installed lightweight \gls{ai} models, avoiding unnecessary data transmissions. However, this also indicates the need for further investigation of the coexistence between push- and pull-based access and the design of efficient \gls{mac} structures for goal-oriented tasks, such as learning and inference. While countermeasures for handling the excess cost of communication and processing before deployment at the devices, e.g., by exploiting well-known compression techniques such as knowledge distillation, can be enacted, these models should be strategically updated taking into account the available energy, traffic load, and the task-specific performance requirements. Therefore, the underlying \gls{mac} mechanisms need to consider \gls{ai} models and data distribution, as well as the cost of their deployment and exploitation \emph{on-device}, to fully support energy-efficient connectivity.

\subsection{6G integration}
\label{sec:6G_Integration}
A seamless integration of a push-pull approach in future 6G networks necessitates both a higher-layer application able to translate the current goal-oriented requirements and the status of the network into a viable allocation of \emph{network} resources, and advanced \gls{rrc} mechanisms capable of dynamic, real-time \emph{radio} resource allocation based on the chosen \gls{mac} structure. The former can be obtained through the development of \gls{o-ran}-compliant xApps/rApps~\cite{oran-architecture}, while the latter requires flexible management of the \gls{phy} channels.

\paragraph{O-RAN-Compliant xApps/rApps}
xApps operating within the \glsentrylong{near-rt-ric} can serve as proxies between the entity requesting the data, e.g., the \gls{dt}, and the \gls{gNB}. xApps can \emph{interpret high-level requirements} arriving through the Y1 interface -- such as data content preferences and deadlines -- and combine this information with \emph{network status feedback} provided by the \gls{gNB} via the E2 interface~\cite{oran-architecture}. This enables the conversion of goal-oriented requirements into resource configurations, to be implemented through dynamic \gls{rrc}. Moreover, xApps can \emph{process data} acquired from the network before the transmission to the end destination via the Y1 interface. Finally, in scenarios where multiple network agents potentially have conflicting objectives, an xApp must aggregate and prioritize these requirements to ensure efficient resource allocations. \minor{This allocation problem is highly dependent on the scenario and the needs of the involved applications, requiring a complex multi-objective optimization which is in itself an open challenge.} When near-real-time is not required, similar operations can be implemented in rApps within the \glsentrylong{non-rt-ric}~\cite{oran-architecture}. 

\paragraph{Dynamic \gls{rrc}}
To effectively implement the resource configurations determined by xApps, the \gls{rrc} must dynamically manage radio resource allocation for each \gls{phy} channel, organizing their placement within the \gls{mac} frame. In 5G, access is performed over the \gls{prach}. This channel can support contention-based transmission in push-pull \gls{mac} designs, i.e., for the push portion of the \gls{cff_frame} frame and throughout the \gls{rcs_frame} frame. Conversely, the \gls{pusch} handles scheduled transmissions, enabling uni-cast pull-based communications within the \gls{cff_frame}. Moreover, the \gls{pdcch} can convey control signaling regarding dynamic \gls{rrc} information, including \gls{mac} frame design  and scheduling directives. A flexible allocation of these channels across available \gls{phy} resources is crucial to accommodate diverse push-pull paradigms. Moreover, integrating energy-efficient technologies such as \gls{wur} introduces additional complexity in control signaling interactions. For example, to enable the \gls{cff_frame} frame with \glspl{ucwu}, a \gls{wus} containing the device ID could be sent via the \gls{pdcch} -- or a newly defined physical \gls{wus} channel -- before each pull slot employing the \gls{pusch}. Then, the \gls{prach} would implement push slots, potentially occurring with a periodicity lower than $10$~ms -- the minimum allowed by the 5G standard -- depending on the frame duration~\cite{cavallero2024co-existence}. Thus, exploring more flexible \gls{rrc} approaches is essential for integrating push-pull coexistence into future 6G networks.

\section{Conclusion}
This work provides \gls{mac} protocol design guidelines based on novel pull- and push-based communication schemes for goal-oriented data transmission in 6G systems. We articulate the need to consider timing and data availability as fundamental measures for assessing data relevance and enabling intelligent network operations, such as adaptive and flexible frame structure to support application-level requirements and \glspl{kpi}. Using examples of goal-oriented scenarios, we present key trade-offs in the coexistence of pull- and push-based communication and postulate a design space for access schemes that includes an additional semantic dimension. We conclude by outlining potential integration challenges and opportunities with the presented push-pull data transmission schemes, particularly when data-driven intelligence and requirements for goal-oriented communication emerge in open networking architectures, such as \gls{o-ran}.   

\bibliographystyle{IEEEtran}
\bibliography{IEEEabrv,ref}
\section*{Biographies}
\begin{IEEEbiographynophoto}{Shashi Raj Pandey} is currently an assistant professor at the Connectivity Section, Aalborg University, Denmark. He received his Ph.D. in computer science and engineering from Kyung Hee University, South Korea, in 2021. His research interests include network economics, wireless communications and networking, distributed machine learning, and semantic communications, and is currently an Associate Editor of IEEE TMLCN.
\end{IEEEbiographynophoto}
\begin{IEEEbiographynophoto}{Fabio Saggese} is currently a  Marie Sk{\l}odowska-Curie Postdoctoral Fellow at the Department of Information Engineering, University of Pisa, where he received his Ph.D. in 2022. His research focuses on PHY/MAC wireless connectivity, goal-oriented communications, and integrated sensing and communications.
\end{IEEEbiographynophoto}
\begin{IEEEbiographynophoto}{Junya Shiraishi} is currently a Marie Sk{\l}odowska-Curie Postdoctoral Fellow at the Connectivity Section, Aalborg University, Denmark. He received his Ph.D. in engineering from Kansai University, Japan, in 2023. His research interests include energy-efficient PHY/MAC protocol designs for IoT networks, tiny machine learning for communication, and goal-oriented communication. 
\end{IEEEbiographynophoto}
\begin{IEEEbiographynophoto}{Federico Chiariotti} is an assistant professor at the University of Padova, Italy, where he also received his PhD in information engineering in 2019. He has authored over 100 peer-reviewed papers on latency optimization and semantic and goal-oriented communication and is currently an Associate Editor of IEEE TWC and IEEE Net. Lett.
\end{IEEEbiographynophoto}
\begin{IEEEbiographynophoto}{Petar Popovski}
is a Professor at Aalborg University, where he heads the section on Connectivity and a Visiting Excellence Chair at the University of Bremen. He is the Editor-in-Chief of IEEE JSAC. His research interests are in communication theory and wireless connectivity.
\end{IEEEbiographynophoto}

\end{document}